\documentclass[aps,prd,twocolumn,superscriptaddress,groupedaddress]{revtex4}  

\usepackage{graphicx}
\usepackage{dcolumn}
\usepackage{bm}
\usepackage{color}
\usepackage{url}
\usepackage{hyperref}

\newcommand{\SNR}{S/N}
\newcommand{\params}{\mathbf{\theta}}
\newcommand{\hof}{\tilde{h}(f;\params)}

\newcommand{\dt}{\delta t}
\newcommand{\pdt}{p(\dt | d_i)}
\newcommand{\jointpdt}{p(\dt | \mathbf{d})}

\newcommand{\N}{{25}}
\newcommand{\stmo}{{${\sim}0.2$ s}}
\newcommand{\styr}{{${\lesssim}0.1$ s}}

\begin{document}

\title{Gravitational Wave Sources as Timing References for LISA Data}

\author{Tyson B. Littenberg}
\affiliation{NASA Marshall Space Flight Center, Huntsville, AL 35812, USA}

\date{\today}

\begin{abstract}
In the mHz gravitational-wave band, galactic ultra-compact binaries (UCBs) are continuous sources emitting at near-constant frequency. 
The signals from many of these galactic binaries will be sufficiently strong to be detectable by the \emph{Laser Interferometer Space Antenna} (LISA) after ${\sim}\mathcal{O}(1\ \text{week})$ of observing.
In addition to their astrophysical value, these UCBs can be used to monitor the data quality of the observatory. 
This paper demonstrates the capabilities of galactic UCBs to be used as calibration sources for LISA by demanding signal coherence between adjacent week-long data segments separated by a gap in time of \emph{a priori} unknown duration.
A parameter for the gap duration is added to the UCB waveform model and used in a Markov-chain Monte Carlo algorithm simultaneously fitting for the astrophysical source parameters.
Results from measurements of several UCBs are combined to produce a joint posterior on the gap duration.
The measurement accuracy's dependence on how much is known about the UCBs through prior observing, and seasonal variations due to the LISA orbital motion, is quantified. 
The duration of data gaps in a two-week segment of data can be constrained to within \stmo\ using {$\mathcal{O}(10)$} UCBs after one month of observing. 
The timing accuracy from UCBs improves to \styr\ after 1 year of mission operations.
These results are robust to within a factor of ${\sim}2$ when taking into account seasonal variations.

\end{abstract}

\maketitle


\section{\label{sec:intro}Introduction}

The \emph{Laser Interferometer Space Antenna} (LISA) will be a unique astronomical observatory, unveiling the gravitational wave sky in the mHz band expected to be richly populated with sources ranging from stellar mass compact binaries in the galaxy, extreme mass ratio in-spirals in galactic centers, to supermassive black hole mergers at high redshifts~\cite{LISA_L3}.

Most expected LISA sources will be long-duration, while the data stream from the satellites will be interrupted due to periodic maintenance and random disturbances.
Many signal classes are identified in the data stream using phase-coherent template waveform models, requiring data segments to be combined at accurate time intervals to prevent biases introduced by artificial phase shifts in the gravitational wave (GW) signals introduced by gap timing errors.
While spacecraft hardware solutions for accurately timing the duration of gaps are understood, persistent and strong gravitational-wave sources can be used as calibration sources to provide an independent measure of start times for data segments.
At the very least, studying the effectiveness of GW sources at measuring gap durations will quantify the error budget for timing accuracy required for unbiased astrophysical parameter estimation.
If the instrumental timing accuracy is superior to what can be achieved using GW sources, the uncertainties in gap duration can be justifiably ignored by the science analysis.  
On the other hand, if hardware solutions provide comparable accuracy to the GW measurement, uncertainties in the gap duration should be incorporated into the science analysis in order to marginalize over the effects on parameter estimation.
This paper will explore how GWs from ultra-compact binaries (UCBs) in the galaxy can be exploited as phase standards to measure the duration of data gaps.

Ultra-compact binaries (UCBs), mostly double white dwarf systems, are prolific gravitational wave sources emitting at near-constant frequency in the LISA band \cite{Hils:1990vc}. 
LISA will detect $\mathcal{O}(10^5)$ UCBs after $\mathcal{O}(1\ {\rm year})$ of observations \cite{Cornish:2017}.
The strongest UCBs will be detectable after ${\sim}1$ week of observations.
As the observing time of the LISA mission increases, the individual UCBs will be increasingly-well measured, improving the accuracy with which the sources can be used to monitor data quality as new data are acquired.

Ref.~\cite{Pollack:2004} studied the impact of data disturbances, including instantaneous drop outs, in the recovery of UCB-like monochromatic signals in a LISA-like data stream and demonstrated success in removing disturbances, although frequent (${\sim}$daily) disruptions were problematic.
The effect of data gaps on parameter estimation of UCBs was studied in Ref.~\cite{Carre:2010} where it was confirmed that minimizing the number of gaps, as opposed to the duration of each gap, results in the least degradation to the measurement of the astrophysical parameters of UCBs.
The study assumed perfect knowledge of the gap start and stop times.

This paper will quantitatively demonstrate how well the brightest UCBs serve as timing standards by demanding coherence across gaps between two short-duration (one week) data segments after one, two, six, and twelve months of prior observing.
The results show that, after one month of observations, the \N\ brightest binaries provide constraints on the duration of gaps in data to better than \stmo\, improving to \styr\ after one year of observing.

Incorporating data characterization parameters into the model for astrophysical sources as a way of characterizing gravitational wave data is not a new idea. 
Using gravitational wave sources to validate ground-based gravitational wave data calibration has been proposed by Ref.~\cite{Pitkin:2016} in the case where an ensemble of joint gamma-ray burst and binary neutron star observations are available.
For a more immediate application, calibration uncertainty has been identified as an additional source of measurement error~\cite{Vitale:2011wu}.
The method described in Ref~\cite{LVC_Calibration} is routinely used to marginalize over calibration uncertainty in the analysis of compact binary mergers observed by ground based detectors (e.g.~\cite{GW150914_PE}). This paper applies a similar conceptual approach to space-based gravitational wave data.

The paper is organized as follows:  Section~\ref{sec:model} describes the model for the LISA observatory and the UCB signals, and modifications made to include the gap parameters. Section~\ref{sec:analysis} describes the analysis, including details about the data and source simulations in Section~\ref{sec:simulations} and the numerical results in Section~\ref{sec:results}. Section~\ref{sec:discussion} summarizes the key results, addresses potential weaknesses in this study, argues why they are permissible, and outlines future directions of research based on these findings.
\section{\label{sec:model} Data Model}

LISA will be a constellation of three free-flying spacecraft, each on independent heliocentric orbits approximately one AU from the sun.
Each is equipped with an optical metrology system to precisely monitor the distances between free-falling test masses housed within the spacecraft.
From the inter-spacecraft ranging measurements, two Michelson-like interferometry signals are synthesized digitally to simultaneously measure the two GW polarizations, while a third Sagnac channel is insensitive to GWs (at low frequencies) and will be used for detector characterization.
The simulated detector data in this study is consistent with the instrument noise levels quoted in the LISA L3 Mission Proposal~\cite{LISA_L3}.
This analysis takes advantage of the noise-orthogonal $A$ and $E$ data streams constructed from linear combinations of the Michelson-like time delay interferometry channels~\cite{Tinto:2002de}.
The $A$ and $E$ data streams assume identical noise in each of the six interferometer links of the spacecraft, and so are an idealized case that will not be realized during mission operations, but are suitable for this proof-of-concept study.

Because the UCBs observable by LISA are widely separated, the orbital evolution due to GW emission is small over the mission lifetime, and thus the sources appear in the data stream at near-constant frequency.
Thus, UCBs do not produce the dramatic ``chirp'' waveforms that are customarily thought of in the context of GW observations of compact binaries.
The simple frequency evolution of UCBs is a double-edged sword. 
On one hand it dramatically simplifies the waveform modeling, as beyond leading order post-Newtonian terms are negligible. 
On the other hand, it is the higher-order post-Newtonian terms in the in-spiral (and the full-relativistic effects in the merger, etc.) that encode the information about the constituents of the binary (e.g. masses and spins).
Furthermore, a subset of UCBs will be interacting with one another, and their orbital evolution will be dominated by mass transfer (e.g. AM CVn systems), causing the orbital period to \emph{increase} during LISA observations~\cite{Nelemans:2004}.
As a consequence, masses of (and the distance to) typical UCBs are not directly observable, but rather combined into phenomenological parameters for the GW amplitude $\mathcal{A}$, frequency $f_0$ (defined at some fiducial time $t_0$), and linear frequency evolution $\dot{f}$ which may, in principle, contain relativistic effects \emph{as well as} contributions from mass transfer. 

The remaining observables are two angles defining the sources position on the sky ${\alpha, \delta}$ and three angles describing the source orientation with respect to the observer ${\iota, \psi, \phi_0}$. 
For this work the location and orientation angles are nuisance parameters which are marginalized in this analysis, though they are of paramount importance for multimessenger observations~\cite{Cooray:2003qm,Littenberg:2012vs,Shah:2014,Korol:2017}.
Together, these quantities form a parameter vector $\params$ which describes a single frequency-domain UCB template.
This study uses the fast-slow decomposition originally described in Ref.~\cite{Cornish:2007if} to forward-model the response of the LISA detector $\hof$ to the gravitational wave source.


Using the UCBs  as time standards requires modifying the waveform $\hof$ with a ``calibration parameter'' $\dt$ which determines the duration of the gap between two segments of data.
Without loss of generality, it is assumed that the start and stop time for the first segment of data, $t_0$ and $t_f$, respectively, are perfectly known.
The $\dt$ then parameterizes the absolute start time of the second segment as $t_f+\dt$ (see Eq. \ref{eq:waveform}). 
A uniform prior $p(\dt)$ for the gap duration is used with support between $\pm20$ s. 



\begin{eqnarray}\label{eq:waveform}
\hof &\rightarrow&  \left \{ 
	\begin{array}{ll}
		\tilde{h}(f,t_0;\params)\ \text{for $t_0<t<t_f$} \nonumber \\
		\tilde{h}(f,t_f+\dt;\params)\ \text{for $t>t_f$} \nonumber
	\end{array}
	\right. \nonumber \\
p(\dt) &=& U({\pm20}{\rm \ s})
\end{eqnarray}


\section{\label{sec:analysis} Analysis}
The goal of this work is to establish an understanding of the capability for astrophysical sources to be used as timing standards. 
This capability is demonstrated by analyzing simulated data containing high \SNR\ UCBs using a parallel tempered Markov Chain Monte Carlo (MCMC) pipeline adapted from Ref~\cite{Littenberg:2011zg}, which was specifically designed for galactic binary searches in LISA data.

\subsection{\label{sec:simulations} Simulations and Procedure}
The simulated galactic population of UCBs used for this study is the same as in the first generation of LISA Data Challenges \footnote{\url{https://gitlab.in2p3.fr/stas/MLDC}}. 
The galaxy simulation combines a population of detached white dwarf binaries from~\cite{Toonen:2012} and interacting binaries (e.g. AMCvn-type systems) from~\cite{Nelemans:2005}.
The binaries are distributed randomly in a model of the galaxy and assigned random orientations with respect to the observer.
From ${\sim}1$ to ${\sim}4$ mHz the LISA band will be dominated by the millions of UCBs which blend together to form an irresolvable foreground. 
Above ${\sim}4$ mHz the remaining UCBs are isolated from one another in frequency and have high \SNR\, making them easily detectable.
From the galaxy simulation, binaries are ranked by \SNR\ and the \N\ loudest systems with $f_0>4$ mHz are selected for study. 
There are high \SNR\ binaries at lower frequencies, however they will be contending with overlapping signals and the galactic confusion noise which make for a more complicated analysis owing to the computational cost of determining the confusion noise level,  contending with it's time-varying power due to variations in the orientation of the LISA detector plane induced by the satellites' orbital motion, and potential covariances with sources nearby in frequency to the binary in question.  
In general, exploiting UCBs for data quality purposes will be most straightforward using isolated binaries at high frequency.

This study does not include binaries already discovered through electromagnetic observations known to be LISA sources~\cite{Stroeer:2006rx,Korol:2017, Kupfer:2018}.  While often discussed as potential calibration sources for LISA--they are commonly referred to as ``verification binaries''--population synthesis simulations predict the existence of significantly stronger GW sources than the verification binaries.  The known LISA binaries are at lower frequency where the noise characteristics are more complicated, due to the dense population of UCBs at those frequencies, meaning longer integration times will be required for these binaries to be detectable.
The high frequency binaries observable by LISA are easier to detect and characterize, and therefore will serve as better phase standards over shorter integration times.

\begin{figure}
	\includegraphics[width=0.5\textwidth]{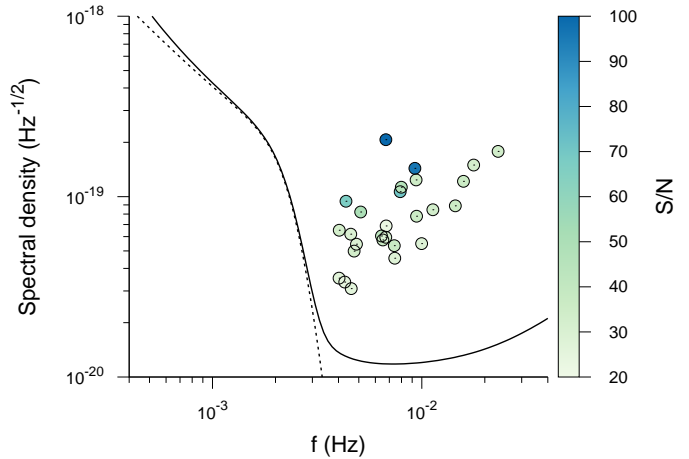}
	\caption{\small Circles show locations of the UCBs in frequency-amplitude space colored by their \SNR\ after two weeks of observing. The black line is the LISA sensitivity curve, including the contribution from the unresolved foreground of low-frequency UCBs in the galaxy (dashed line). The foreground is the residual signal from UCBs after all resolvable signals have been identified and subtracted. This example is from the expected residual after one year of LISA operations. For this study,  only binaries with frequencies above the confusion noise were selected to avoid complexities due to seasonal variations of the foreground and source overlap.}
	\label{fig:sensitivity}
\end{figure}

Figure~\ref{fig:sensitivity} shows the strain amplitude of the simulated sources as a function of their GW frequency, compared to the LISA sensitivity curve and an estimate of the confusion noise level after one year of observations. 
The sources are colored by their \SNR\ which depends on the location and orientation and so can not be simply read off from the height above the sensitivity curve.

The parameter $\dt$ introduced to measure gap duration is correlated with the frequency parameter $f_0$ of each UCB signal. 
To improve the measurement of $\dt$, results from several binaries are combined, and the knowledge about each source that builds up over time is incorporated into the analysis.
To put it differently, when analyzing short segments of data, priors on the source parameters constructed from earlier observing times are enforced. 
While the degeneracies between $\dt$ and source parameters are not broken, constraints on the astrophysical parameters which build up over time will propagate through the analysis to improve the determination of $\dt$ in newly acquired data.
If this procedure were to be adopted during mission operations, the analysis would need to be diligent about which data were used to fit gap durations and which were used to build priors to avoid ``double-counting'' the data i.e., using the same segments for analysis as were used to build priors.

For this study, separate priors are constructed after one month, three months, six months, and one year of prior observations for each high \SNR\ UCB. 
From the posteriors inferred from these ``pilot'' MCMC runs, uniform priors are constructed for the eight source parameters spanning the $90\%$ credible intervals of the marginalized posteriors.
Distilling the full posteriors to a set of 1-dimensional uniform distributions is overly conservative, and it ignores valuable information about the source parameters which, when included, would make the priors more representative of what is known about each binary.
However, for this proof-of-principal exploratory study, this short cut is sufficient considering that at this time the actual scenario of the instrumental uncertainties in gap frequency and duration are yet to be determined.
Developing efficient ways of building priors from previous observations, and updating posteriors as more data are acquired, will be an important development for LISA data analysis given the long-duration signals which will be observed.

To demonstrate the $\dt$ measurement capabilities, two weeks of data are simulated as two separate one-week segments.
The data are analyzed demanding signal coherence in the two segments, fitting for the start time of the second segment relative to the end of the first along with the source parameters. 
This is done independently for each of the \N\ brightest UCBs to determine the marginalized posterior distribution of $\dt$ for each injection $i$
\begin{equation}
\pdt = \int  p(\dt,\params | d_i)\ d\params
\end{equation}
Because $\dt$ is common to each data segment, analysis of each binary is an independent measure of the gap duration. 
Thus the joint posterior on $\dt$ is approximately the product of the marginalized distributions $p(\dt | \mathbf{d})=\prod_i \pdt$ (ignoring covariances between UCB parameters introduced by each sharing the $\dt$ parameter).
The marginalized posteriors for $\dt$ for the high \SNR\ binaries are, to a good approximation, Gaussians. For simplicity, each $\pdt$ is fit to a normal distribution so that the product can be computed analytically. 
More accurate approaches would be to use a hierarchical analysis to characterize the posterior on $\dt$ (e.g., see Ref.~\cite{Mandel:2010} for a GW-centric example) or, better yet, to perform a global fit to the \N\ UCBs with $\dt$ being common for each source, but such complexity is not needed for this exploratory study.


For one-week data segments the strength of the signal, and therefore it's utility as a calibration source, will depend on the source's position relative to the plane of the LISA constellation, causing the measurement accuracy of the source, and therefore the calibration parameter, to vary over the course of each year. 
To account for the dependence on the location of LISA in its orbit, the analysis is repeated once during each quarter of the year to bracket  the range of outcomes for the galaxy realization and mission configuration used in this study.  

\subsection{\label{sec:results} Results}

\begin{figure}
      \includegraphics[angle=270,width=0.5\textwidth]{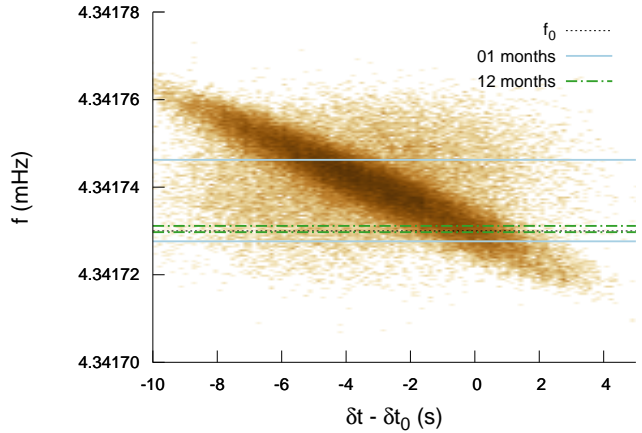}
    \caption{\small Two-dimensional marginalized posterior distribution function for $f_0$ and gap duration $\dt$ (shifted so that 0 is the true value) from analyzing two weeks of data without assuming any prior knowledge of the source, demonstrating the covariance between source frequency $f_0$ and gap duration.  Horizontal lines enclose the 90\% credible region after one month (light blue, solid line) and twelve months (dark green, dot-dashed line) of prior observing.  The black dotted line is the true value of $f_0$ for the simulated source. Priors constructed from pilot runs on UCBs will propagate to improved constraints on the gap duration.}
  \label{fig:covariances}
\end{figure}

First covariances in a single source between the gap duration parameter $\dt$ and $f_0$ are investigated, to motivate the importance of building up priors from the LISA observing time in order to exploit UCBs as calibration sources in short segments of data. 
Correlations with other phase parameters of the UCB were not found to be important, as $\dot{f}$ and $\phi_0$ are poorly constrained when analyzing  short data segments.
Figure~\ref{fig:covariances} shows the marginalized two-dimensional posterior distribution functions between $\dt$ and $f_0$.  The black dashed line is the true value for the simulated source, while the colored horizontal lines show the 90\% credible intervals of the source parameters after one (light blue, solid lines) and twelve (dark green, dot-dashed lines) months of prior observation.
As is evident in the correlations between the source and gap duration parameters, constraints on $\dt$ will rely on information about the sources which will build up over the mission lifetime. Note that for prior observation times of greater than one year, the improvement in measuring $\dt$ saturates i.e., further restricting the prior range on $f_0$ does not notably restrict $p(\dt|d)$, as it is fundamentally limited by the short-duration segments being analyzed.

\begin{figure}
      \includegraphics[angle=270,width=0.5\textwidth]{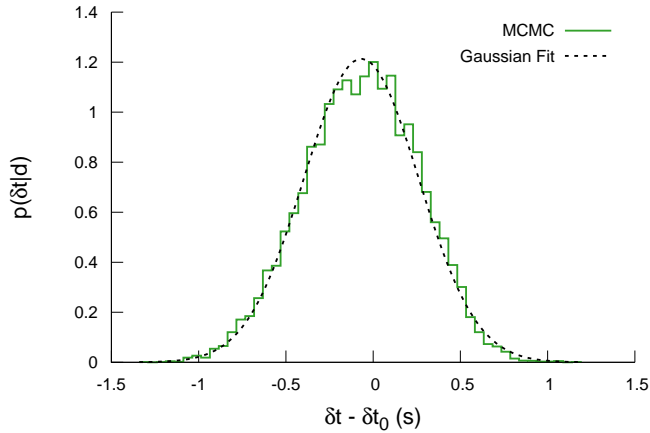}
    \caption{\small Example Gaussian fit (black, dashed line) to the marginalized posterior distribution function for $\dt$ (green, solid histogram). The posterior is shifted in time so that the true value is at 0. Fitting the posterior to a Gaussian is a satisfactory approximation for this study. This demonstration used a ${\sim}7$~mHz source after two months of prior observing.}
  \label{fig:gaussian_test}
\end{figure}

After adopting uniform priors over the 90\% credible intervals from the pilot runs, each of the \N\ brightest binaries are independently analyzed and the marginalized posteriors on $\dt$ are fit to a Gaussian distribution. 
Using uniform priors over the 90\% credible intervals of the pilot runs, and multiplying Gaussian fits to the $\dt$ posteriors, is a simplified approach to what  would be deployed for a production-quality version of this analysis, but is sufficient for exploring the potential for UCBs to function as phase standards in LISA data.
Figure~\ref{fig:gaussian_test} shows a representative marginalized posterior on $\dt$ from the MCMC along with the Gaussian fit, demonstrating the consistency between the sampled distribution and the Gaussian approximation.

\begin{figure*}
      \includegraphics[width=\textwidth]{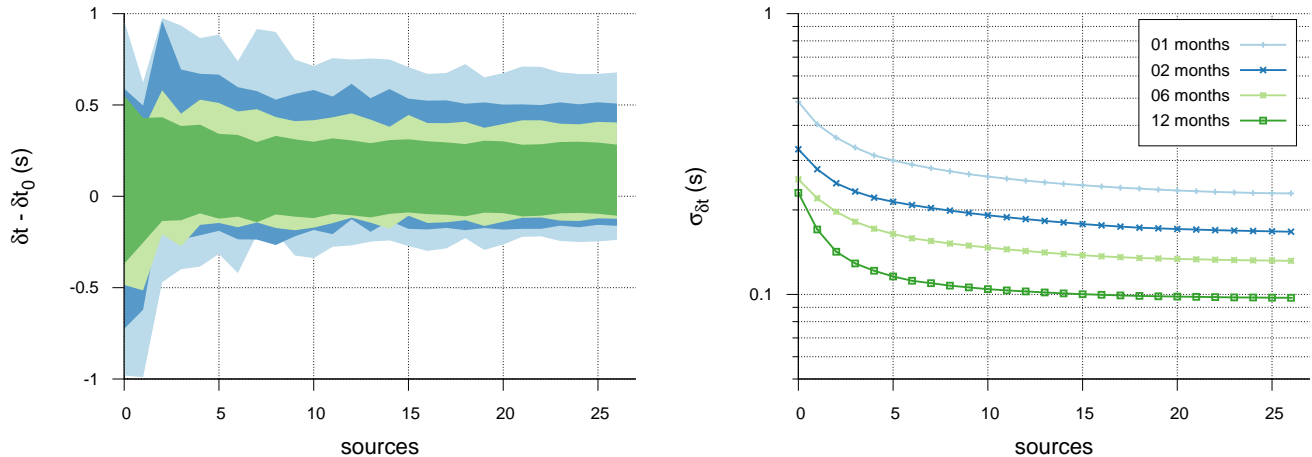}
    \caption{\small Left panel: $2\sigma$ intervals of the joint posterior on the gap duration $\dt$ separating two 1-week segments of data as a function of the number of sources in the fit.  The joint posterior $\jointpdt$ converges towards the true value of $\dt - \dt_0=0$ as the number of sources increases. Different color bands correspond to priors built from one month (light blue) two months (dark blue), six months (light green) and twelve months (dark green) of observing. Right panel: The same results now showing the standard deviation $\sigma_{\dt}$ of the joint posterior as a function of the number of sources, reaching 0.2 s (0.1 s) for 1 month (12 months) of prior observing.}
  \label{fig:dt}
\end{figure*}

From the analytic fits to each $\pdt$, the joint distribution on the gap duration from the ensemble of high-frequency, high \SNR\, binaries is computed as the product of the Gaussian fits to the individual posteriors. 
The left panel of Figure~\ref{fig:dt} shows the $2\sigma$ envelopes for the joint posterior as a function of how many binaries are used in the fit.  
The four bands correspond to the joint posteriors using the four different prior observing scenarios in this analysis:  One month (light blue), two months (dark blue), six months (light green), and twelve months (dark green).
From this figure it is apparent that the posterior is converging towards the true value at $\delta t - \delta t_0 = 0$.
The right panel, using the same color scheme, quantifies how the standard deviation ($\sigma$) of the joint distribution depends on the number of binaries in the fit, reaching a value between ${\sim}0.2$ and ${\sim}0.1$ when 20 binaries are employed in the joint fit, depending on the prior observing time. 
Note that, should the LISA science data be sampled at or below 1 Hz, the UCB measurement provides sub-sample accuracy on its measurement of the gap duration.

\begin{figure*}
      \includegraphics[angle=270,width=\textwidth]{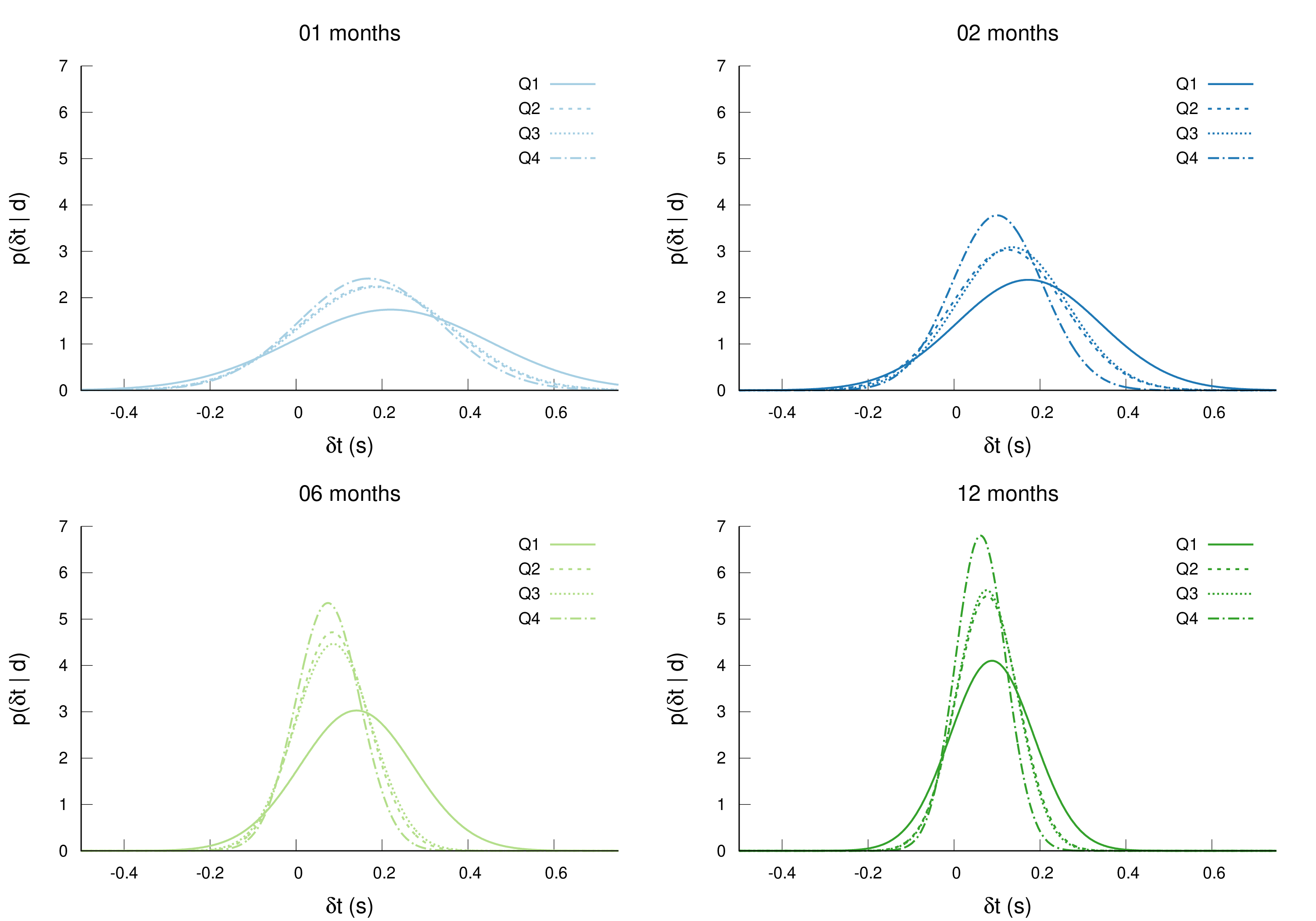}
    \caption{\small Demonstration of the seasonal variation in the joint posterior of $\dt$ using the \N\ brightest UCBs after one (top left), two (top right), six (bottom left), and twelve (bottom right) months of prior observing.  The different distributions are the results from simulating data during each quarter of the LISA constellation's yearly orbit. While the seasonal variation is evident, the constraints on $\dt$ are of similar scale.}
  \label{fig:quad}
\end{figure*}

Because of the short duration data segments used to fit for the gap duration, constraints derived from the analysis of UCBs will exhibit seasonal variation as the orientation of the LISA constellation changes with respect to the source during the year. 
To test the impact of these seasonal variations, the analysis is repeated simulating two week segments of data from each quarter of the year and the joint posteriors on $\dt$ are compared after combining the top \N\ binaries in the catalog.
Figure~\ref{fig:quad} shows the joint posterior $\jointpdt$ for the different quarters of the year.
While the variation is apparent in the results, the fidelity of the $\dt$ constraints are robust to within a factor of ${\sim}2$ throughout the LISA orbit.

\section{\label{sec:discussion} Discussion}
This paper investigated how bright UCBs could be used as phase standards in LISA data to constrain the duration of data gaps.
This work demonstrates that UCBs can be used as calibration sources on short segments of data.
The duration of a data gap in a two-week segment of data can be constrained to within \stmo\ using {$\mathcal{O}(10)$} UCBs after one month of observing. 
The timing accuracy from UCBs improves to \styr\ after {1 year} of observing.
This result is robust to seasonal variations as the orientation of the LISA constellation changes with respect to the GW sources.

The results of this study are a useful benchmark for timing accuracy of gap durations for LISA, as it concerns UCBs.  
If gap durations are constrained through spacecraft ``house-keeping'' data to much better precision than found here, the effect will be negligible in the analysis of UCBs.
Alternatively, should there be some ambiguity in the duration of a data gap to larger than the amounts demonstrated here, UCBs can be used to better-constrain the absolute start times of data segments.

Future investigations aiming to measure the impact of data gaps on LISA observations should consider how the uncertainties in gap duration as constrained by UCBs influence the characterization of other astrophysical sources, such as supermassive binary black hole mergers, to determine the conditions under which the level of timing accuracy demonstrated here is sufficient for other sources.

The analysis reported here is a proof-of-concept demonstration using methods similar to production-level algorithms for GW data analysis~\cite{LALInference,Littenberg:2011zg, LVC_Calibration}.
That being said, there are some places where the numerical experiment performed here would be vulnerable due to simplifications made in this study.

The procedure of constructing uniform priors over the 90\% credible intervals from marginalized posteriors of the pilot runs is overly simplistic. 
Information obtained from the pilot runs, including the shape of the distributions and covariances between parameters, is lost by this approach. 
More egregiously, there is the possibility that the true source parameters will be excluded by the overly-simplified priors.
It is not expected that addressing these issues would lead to qualitatively different conclusions about the UCBs' utility in constraining the duration of data gaps, as the analyses on the week-long segments of data basically saturate their priors. 
To put it differently, small changes to the priors will lead to similarly small differences in the posteriors because the likelihood on the short segments of data is nearly constant over the priors from the pilot runs. 
In a production-quality analysis these details will need to be addressed even if their effect is small.
To that end, investigations into strategies to build joint priors on the astrophysical parameters from posteriors inferred by pilot runs, which are more representative of the full $N$-dimensional distributions, while still being efficiently implementable in stochastic sampling algorithms, have begun.

Another weakness of this work is its dependence on a single realization of the UCB galactic population.
The top \N\ binaries out of a population of ${\sim}10^4$ resolvable sources are in the tail of the source distribution, and characteristics of this sub-population will have a large variance.
However, high frequency binaries in the galaxy are generically high \SNR\ sources, and while this study used \N\ binaries, the constraints on $\dt$ begin saturating after ${\sim}5$ to ${\sim}15$ binaries, depending on the prior observation time.
A more thorough investigation into the variability of the LISA UCB sources, including into the population of high frequency detached binaries which are of scientific interest beyond their utility as calibration sources, as a result of different population simulations, is an area of ongoing study. 

Finally, the dependence on the prior observations of binaries to confine source parameters when analyzing the short data segments raises the question of how those priors are obtained.
The pilot runs to establish the priors did not consider that those data will also contain gaps.
Therefore, the procedure demonstrated here assumes a reliable gap measurement from monitoring the housekeeping data during the first several months of the mission operations. 
In its absence, assuming uncertainties in the gap durations from the instrumental measurements are dominated by statistical error as opposed to a systematic offset,  the performance demonstrated here would still be achievable, but the convergence time, i.e. the amount of prior observing needed for establishing similarly precise priors, would increase.
A study incorporating such a level of detail should be undertaken as the LISA mission design reaches maturity, leading to more realistic scenarios for the acquisition of data and the uncertainty budget in the data gaps. 

This work demonstrates that UCBs can be used to measure the duration of data gaps and, as a corollary, places an upper limit on the uncertainty of the instrumentally-measured gap durations for it to have a negligible effect on UCB measurement.
This paper also provides guidance for how analysis of LISA data will interact with the uncertainty budget of the observatory, and how UCBs can provide, at a minimum, a reserve means for maintaining phase-coherence in the LISA data stream.

\section{Acknowledgments}

The author would like to acknowledge contributions to the analysis software used here by Neil Cornish and Travis Robson.
This work was supported by NASA grant NNH15ZDA001N-APRA.
The author would also like to acknowledge the support of the NASA LISA Study Office.

\end{document}